# Enhanced interfacial thermal conductance in functionalized Boron Nitride/Polylactic acid nanocomposite: A molecular dynamics study

G. Jamirad[1], A. Montazeri[1,3], A. Rajabpour[2,3,*]

[1]Computational Nanomaterials Lab (CNL), Faculty of Materials Science and Engineering, K. N. Toosi University of Technology, Tehran, Iran
[2]Advanced Simulation and Computing Laboratory (ASCL), Mechanical Engineering Department, Imam Khomeini International University, Qazvin, Iran
[3]School of Nano Science, Institute for Research in Fundamental Sciences, IPM, Tehran, Iran

## Abstract

The relatively low thermal conductivity of biodegradable polylactic acid (PLA) has limited its applications in various fields. To address this issue, the incorporation of nanofillers, such as boron nitride nanosheets (BNNSs), has emerged as an effective method to enhance PLA's thermal properties. However, the thermal conduction of polymer-based nanocomposites is strongly influenced by interfacial thermal resistance. In this study, we investigate the impact of pristine and surface-treated BNNSs on the thermal behavior of PLA using molecular dynamics simulations. To enhance interfacial interactions and reduce chain mobility during heat transfer, we chemically modify the surface of BNNSs by introducing three different functional groups ($NH_2$, OH, and COOH) with varying polarities. Our findings suggest that oxygen-containing groups, namely –OH and –COOH, exhibit stronger interfacial interactions compared to the other cases. We also systematically apply different percentages of these functional groups (i.e., 2.5, 5, and 7.5%) and observe that a higher number of functional groups leads to a greater improvement in interfacial thermal transport, attributed to the enhanced phonon coupling effect. To complete the discussion, we thoroughly study the influence of random and agglomerated patterns of functional groups distribution.

* Corresponding author: rajabpour@eng.ikiu.ac.ir (A. Rajabpour)

## 1. Introduction

In recent years, there has been a growing interest in the development of polymer matrix nanocomposites (PMNCs) among scientific and industrial communities, due to their attractive physical and mechanical characteristics. Green PMNCs, in particular, have drawn considerable attention owing to their unique properties, such as lightness, high heat capacity, desirable chemical resistance, stability, and eventually, biological characteristics, such as biocompatibility and biodegradability [1, 2]. Polylactic acid (PLA) has proven potential to replace conventional polymers for industrial purposes (e.g., textiles, electronic devices, and packaging) or serve as a leading biomaterial in a wide range of biomedical applications like absorbable surgical fibers, orthopedic, and drug delivery devices [3, 4]. This is due to the outstanding features of PLA, including high strength and Young's modulus, biodegradability, non-immunogenicity, and ease of processing [5–7]. However, its extensive usage as a thermoplastic can be overshadowed by its low thermal conductivity resulting from phonon scattering [3, 8].

To enhance the thermomechanical properties of polymers, particularly their thermal conductivity, introducing nanofillers (e.g., nanoparticles, nanosheets, and nanotubes) to their matrices have proven to be an effective tool [9–11]. Among these, boron nitride nanosheets (BNNSs), which are referred to as "white graphene" due to the similarity in their honeycomb lattice structures, have been widely utilized in reinforcing various polymer matrices such as PLA. This arises because in addition to the superior mechanical properties and thermal conductivity, BNNS has been recognized to be more electrically insulated, more thermally and chemically stable, and better transparent against visible light compared to graphene [12–14]. Furthermore, PLA-based NCs embedded with BNNSs are ideal candidates for biomedical applications, including bone and dental implants owing to the biocompatible nature of these nanofillers [15–17]. To develop heat transfer models in the human body containing implant devices, as well as to protect the surrounding tissues and cellular functions, a thorough understanding of the thermophysical characteristics of such biomaterials is crucial to properly cool the device, as a relatively small surface area has to dissipate the generated heat [18–21]. It is worth mentioning that BNNSs have an extremely high thermal conductivity of around 600 W/mK, making them suitable for designing highly thermally conductive NCs [22–24]. In this context, Kong et al. [25] prepared a cellulose/BNNS composite film, which demonstrated an enhanced thermal conductivity of 9.5% compared to that of a pure cellulose film. Using polyimide (PI) composite films containing 7 wt.% BNNSs, Wang et al. [26]

achieved an in-plane thermal conductivity of 2.95 W/mK, which is 1080% higher than that of the untreated PI. Furthermore, PMMA nanocomposites infused with BNNSs demonstrated a thermal conductivity of 2.6 W/mK, showing an increase of 17-fold in comparison with the neat polymer matrix [27]. Similarly, the introduction of only 4 wt.% BNNS enhanced the thermal conductivity of PLA from 0.25 to 0.30 W/mK, leading to an improvement of 21.4% [23]. However, the interfacial area can limit the ability of nanocomposites to improve their thermal conductivity [28, 29].

As discussed extensively in the literature, heat management in newly developed NCs is primarily influenced by interfacial thermal conductance. Due to phonon scattering at the interface, heat loss is a common phenomenon during the thermal energy transport across two distinct phases of nanocomposites. To address this issue, interfacial thermal resistance (ITR) or Kapitza resistance is used to quantitatively analyze the hindrance of heat transfer at the interfacial area of PMNCs [30, 31]. It should be noted that several procedures have been utilized to reduce the ITR of these nanocomposites, including surface functionalization, ordering alignment, and bridging of thermally conductive nanofillers [32–37]. Shi et al. [38] revealed that decorating hydroxyl (–OH) groups on BNNSs embedded in poly(vinyl alcohol) (PVA) could greatly enhance the thermal conductivity of composite films. Similarly, Liu et al. [39] investigated in detail the effect of BNNS functionalization with 3-aminopropyltriethoxysilane group on the improvement of thermal conductivity of epoxy-based NCs. The discrepancy between the performance of functionalized and non-functionalized nanosheets was ascribed to the better interface cohesion and dispersion of BNNSs in the cases embedded with surface-treated nanosheets, leading to a decrease in nanocomposites ITR via reduced phonon scattering.

In recent years, many studies have been conducted to explore the thermal resistance of nanoscale materials using both experimental techniques and computational simulations at the atomic scale. However, experimental-based methods are time-consuming and demanding when studying nanoscale interactions at interfaces across a wide range of fields [40–42]. Therefore, molecular dynamics (MD) simulation has become a renowned technique for exploring nanoscale systems, particularly their Kapitza resistance. Among different techniques, transient non-equilibrium MD (NEMD) simulations are inspired by the experimental measurements of the ITR [43]. Based on the above discussion, utilizing nanostructures with pristine and functionalized surfaces has proven to be an effective method for enhancing the thermal properties of polymer matrices [9–11, 22–27].

However, no atomic-scale studies have addressed the role of interfacial area on the thermal transport in BNNS/PLA nanocomposites. Moreover, the effect of nanofiller surface-treatment on the thermal features of this type of PMNCs remains unknown. To provide a profound understanding of this issue, several MD simulations are carried out to examine the ITR of BNNS/PLA nanocomposites in the presence of various functional groups such as –OH, –COOH, and –NH$_2$ at different coverage degrees. This is followed by the evaluation of vibrational density of states (VDOS) vs. frequency and RDF analyses to obtain the origin of the ongoing phenomena at the interface.

## 2. Methodology and simulation details

### 2.1. Nanocomposite construction

The construction of the nanocomposite models began by creating polylactic acid chains using the Materials Studio software [44]. To achieve this, a lactic acid monomer (C$_3$H$_4$O$_2$) was created, which had two excess hydrogen atoms at the head and tail of the sequence. By connecting the hydrogen atoms, a polymer chain with a length of 20 monomers was created (see Fig. 1). In the next step, a cubic simulation cell measuring 6 nm in size was constructed, which consisted of 110 randomly distributed polymer chains to mimic the polymeric model structure. Finally, an initial energy minimization process was performed on the polymeric computational cell using the conjugate gradient (CG) method.

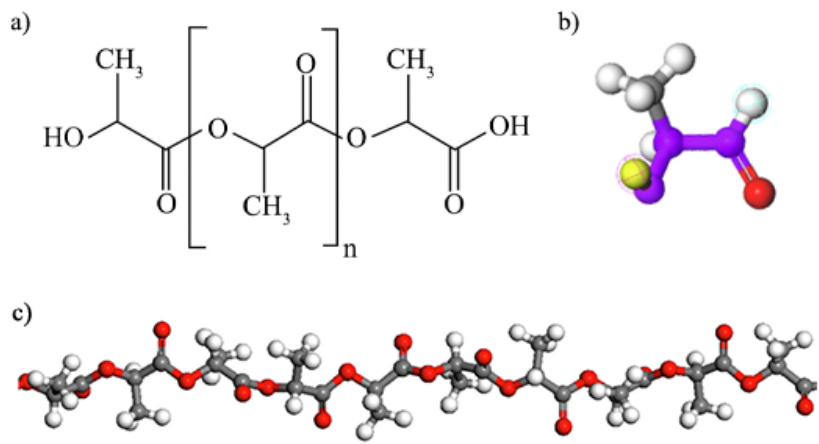

Fig. 1. (a) Chemical structure of the PLA chain, (b) PLA monomer, and (c) PLA chain consisting of 20 monomers created using the Materials Studio software.

To construct the nanocomposite models, polymer chains were randomly distributed and embedded with a BNNS having a dimension of 4.9×2.3 nm$^2$, accounting for 3 wt.% of the composition. It is worth mentioning that hydrogen atoms were added to the edge boron and nitrogen atoms to mitigate spurious non-bonded effects [45]. As part of the investigation into how functionalization affects the thermal properties of the introduced nanocomposite samples, some covalent bonds within the nanosheet structure have been broken to create vacant bonding electrons, enabling chemical functionalization of the BNNS [46–48]. To thoroughly examine this matter, three different functional groups (–COOH, –OH, and –NH$_2$) were selected at coverage degrees of 2.5, 5, and 7.5%. The subsequent step involved additional CG energy minimization and annealing using the open-source LAMMPS MD simulator [49]. Each model experienced equilibration through five loops, employing an annealing time-step of 0.25 fs. Initially, an isothermal-isobaric ensemble (NPT) was assigned at an initial temperature of 298 K for 100 ps, with the temperature and pressure damping parameters of 25 fs and 250 fs, respectively. Subsequently, in a canonical ensemble (NVT) over 25 ps, the temperature was raised from 298 K to 500 K, exceeding the glass transition temperature of PLA. The system then remained at this temperature for an additional 250 ps using the same ensemble. Afterward, the system was cooled back to the room temperature and maintained for another 100 ps under NVT. To guarantee the absence of any residual stress, this procedure was repeated four more times [50, 51]. Ultimately, a density of 1.23 g/cm$^3$ was achieved, which is consistent with both experimental and MD results ranging from 1.20 to 1.25 g/cm$^3$ [50–53].

Figure 2 depicts the annealed nanocomposite model, showcasing the various types of BNNS utilized in this study. As discussed in the literature, several independent factors, such as temperature, pressure, dimensions, polymer cross-linking density, and interface morphology, influence the thermal behavior of nanostructures, particularly the interfacial thermal resistance and thermal conductivity of polymer-based nanocomposites [54–57]. In fact, as the dimensions or system size decrease, thermal transport can behave abnormally. The number and thickness of the nanosheets can also affect the in-plane thermal conductivity of the system, through phonon scattering. However, this study focuses on the interfacial region of the nanocomposite, specifically on the thermal resistance at the interface rather than the thermal conductivity. Therefore, each of the previously discussed variables has been kept constant to enhance intersystem comparisons across all simulations [55, 57].

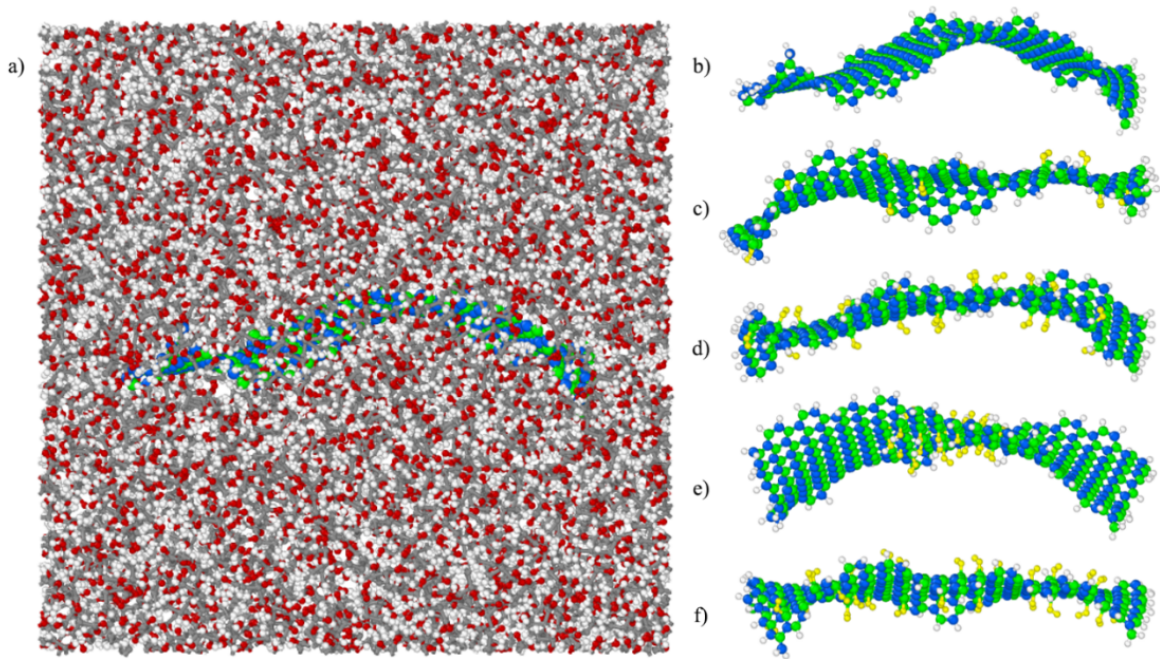

Fig. 2. (a) Annealed PLA embedded with various reinforcements: (b) untreated BNNS, (c) randomly functionalized with 2.5% OH groups, (d) randomly functionalized with 5% OH groups, (e) agglomerated reinforcement with 5% OH groups, and (f) randomly functionalized with 7.5% OH groups.

## 2.2. Details of MD simulation

The present study utilized the LAMMPS package [49] to conduct MD simulations. The Consistent Valence Force Field (CVFF) was chosen to model atomistic interactions in PLA, based on the positive results of previous studies [50, 51]. The CVFF force field accounts for bond stretching, bend angles, internal rotation for dihedral angles, out-of-plane improper angles, and the interplay between these components governing non-bonded interactions (i.e., van der Waals and Coulombic). Although the selected force field produces accurate results, certain coefficients associated with bend angles, which are not explicitly mentioned, must be assumed by a similar force field such as CHARMM. Lange et al. [51] provided a comprehensive explanation of these details. The interactions between boron and nitrogen atoms were modeled using the Tersoff potential, which is well-established in describing the properties of BNNS [58–62]. The parameters for this potential were derived from the study conducted by Matsunaga et al. [63, 64], which were subsequently modified by Mortazavi et al. [59, 65]. Numerous research studies have successfully evaluated the mechanical and thermal properties of polymeric nanocomposites by combining

CVFF with many-body potentials like Tersoff and Airebo [66–68]. As CVFF does not support boron atoms, the Dreiding force field was employed to calculate the non-bonded interactions involving these atoms and the constituents of PLA. The desired parameters were extracted using the Dreiding force field [69]. To account for long-range Coulombic interactions, the particle-particle particle-mesh (PPPM) solver was utilized with a cutoff radius of 12.0 Å. Periodic boundary conditions (PBCs) were applied in all directions, and the integration of Newton's equations of motion was performed using the velocity-Verlet algorithm.

## 3. Results and discussion

### 3.1. Model validation

To validate our computational models, we conducted two standard tests to determine the glass transition temperature (Tg) of PLA and evaluate the tensile behavior of BNNSs. The Tg of amorphous polymers is a key property that can be examined to verify the simulation parameters associated with the polymeric part of the nanocomposite models. It has been demonstrated that amorphous polymers undergo significant changes in their thermal properties, characterized by higher molecular mobility, once they pass their Tg [70]. To determine the Tg of PLA after equilibration, we monitored the specific volume of the polymer models at a constant pressure as a function of temperature. We employed a time-step of 0.4 fs, gradually decreasing the system temperature while applying PBCs in all directions. This method has been successfully utilized in several previous studies [50, 51, 71, 72]. As illustrated in Fig. 3, a reduction in temperature from 500 K to 100 K resulted in an abrupt change in the slope of the curve at 325 K. This turning point, commonly observed in amorphous polymers, represents a significant shift in their mechanical and thermal characteristics and is referred to as the glass transition temperature of the abovementioned polymers [73]. The obtained Tg, as summarized in Table 1, demonstrates excellent agreement with previous experimental and numerical studies, thereby validating the methodology and simulation parameters employed in this work for modeling the PLA matrix.

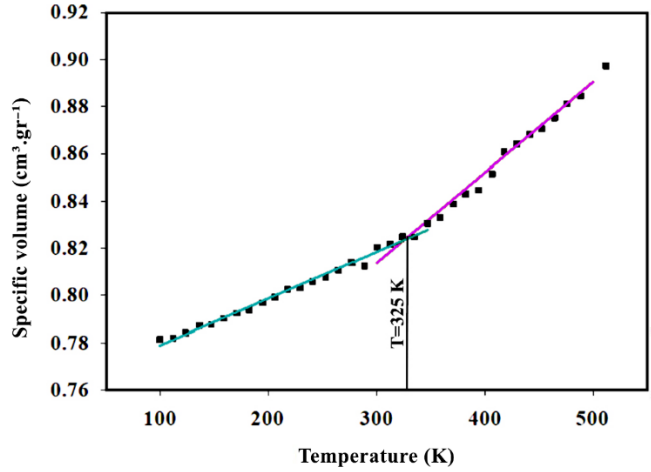

Fig. 3. Calculation of the Tg of the PLA matrix using a plot of specific volume vs. temperature based on MD simulation.

Table 1. Comparison of the determined Tg with those obtained exponentially or numerically in the literature.

| Reference | Approach | Tg (K) |
|---|---|---|
| **Present work** | MD | 325 |
| [74] | Experimental | 327 |
| [75] | Experimental | 328 |
| [76] | Experimental | 337.3 |
| [51] | MD | 345 |
| [50] | MD | 340 |

Moving on to the second validation test, armchair BNNSs were subjected to uniaxial tensile test at a constant strain rate of $1\times10^{-5}$ $ps^{-1}$ after energy minimization using the CG method [62]. PBCs were assigned for the two lateral directions. This process was carried out at the room temperature (i.e., 298 K) by means of the Nose-Hoover thermostat [77, 78]. Figure 4 illustrates the tensile behavior of the BNNS introduced in Section 2.1, providing insights into its mechanical properties such as fracture stress, fracture strain, and Young's modulus. To facilitate comparison, Table 2 presents the results obtained in this work alongside data reported in previous studies. Despite variations in methods and settings, the results generally show agreement, confirming the validity of the simulation details, particularly the employed potential, in accurately modeling the BNNSs.

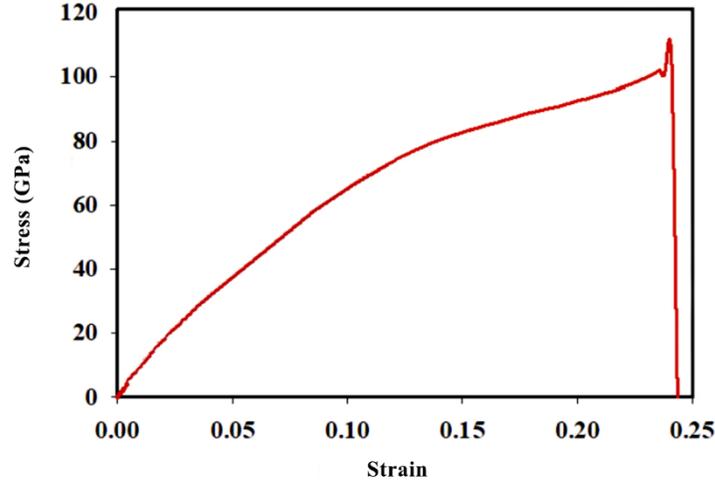

Fig. 4. Stress–strain curve of a 4.9×2.3 nm$^2$ armchair BNNS sample under uniaxial tensile test with a strain rate of $1\times10^{-5}$ ps$^{-1}$ at room temperature (298 K).

Table 2. Tensile test results of the BNNS in the present study in comparison with the available data in the literature.

| Reference | Method | Fracture stress (GPa) | Fracture strain | Young's modulus (GPa) |
|---|---|---|---|---|
| **Present work** | MD (Tersoff) | 118 | 0.24 | 820 |
| [12] | Experimental | 70.5 | - | 865 |
| [62] | MD (Tersoff) | 126 | 0.28 | 740 |
| [59] | MD (Tersoff) | 170 | 0.28 | 850 |
| [79] | DFT | 100 | 0.23 | 996 |

### 3.2. ITR between BNNS and PLA

To gain a comprehensive understanding of the impact of BNNS on the thermal behavior of BNNS/PLA nanocomposites, we employed the transient non-equilibrium molecular dynamics (TNEMD) approach, also known as the pump-probe method [80], to determine the Kapitza resistance in the interfacial region of the samples [81, 82]. This technique offers several advantages over traditional NEMD methods as it focuses on capturing the dynamic thermal response of the system while reducing computational cost [83]. Initially, we conducted a supplementary energy minimization process, followed by applying the NVT ensemble to the simulation cell at an initial temperature of 300K for 100 ps with a time-step of 0.5 fs. Over the next 100 ps, we increased the temperature of the BNNS to 400 K while maintaining the PLA at 300 K using another NVT ensemble. In accordance with Fourier's law of heat transfer, the rise in temperature acts as a driving force for the increase in heat flux [84]. Subsequently, a microcanonical ensemble (NVE) was applied to the embedded nanofiller, allowing heat transfer from the BNNS to the host polymer. The temperature of each constituent was then calculated at each time-step during the simulation, as depicted in Fig. 5. To reduce noise, the obtained temperature and energy values were averaged over every 100 time-step.

The determination of the ITR in nanocomposites involves considering a combination of the energy balance equation and Newton's law of cooling. The former explains energy conservation within a system using thermodynamics and heat transfer, while the latter states that the heat loss rate of a body is proportional to the temperature difference between the body and its surroundings [85]:

$$\dot{Q} = hA(T_{Body} - T_{Enviroment}) \tag{1}$$

Here, $\dot{Q}$ represents the heat transfer rate, h is the heat transfer coefficient, A is the area through which heat transferred, and the last term denotes the aforementioned time-dependent temperature difference. Due to the enclosed 2D nanostructure, the only path for dissipating thermal energy is through the polymer.

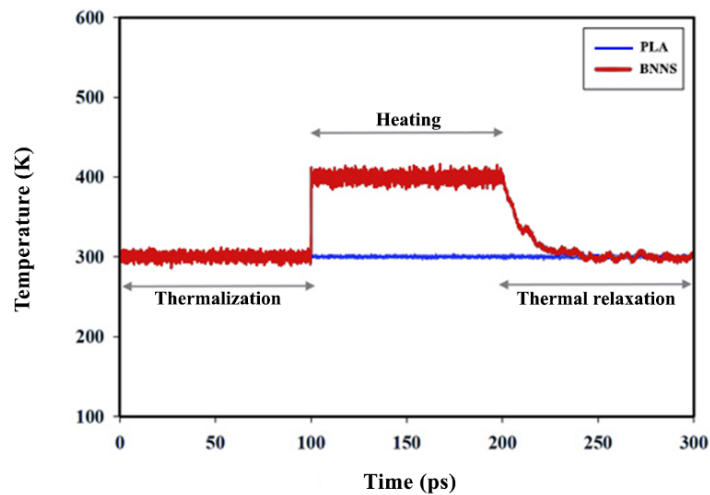

Fig. 5. Temperature profile associated with BNNS relaxation and heat dissipation (red) and PLA temperature (blue) over simulation time.

Applying the first law of thermodynamics to the process gives:

$$\frac{\partial E_t}{\partial t} = \dot{Q} = -\frac{A}{R}(T_{BNNS} - T_{PLA}) \qquad (2)$$

In this equation, $E_t$ denotes the total energy of the BNNS and $R = 1/h$ is the ITR between the two distinct phases of the nanocomposite model. Assuming that $R$ remains constant over time, Eq. (2) can be integrated as follows:

$$E_t = -\frac{A}{R}\int_0^t (T_{BNNS} - T_{PLA})dt \; + E_0 \qquad (3)$$

where, $E_0$ is the initial total energy of the BNNS. By identifying the temperature difference between the constituents of the model at each time step and using the trapezoidal rule for numerical integration, the value of $\int_0^t \Delta T \, dt$ can be calculated during the simulation period. Figure 6 illustrates a linear relationship between the total energy of the BNNS ($E_t$) and $\int_0^t \Delta T \, dt$ , allowing the ITR of the nanocomposite sample to be easily determined from the slope of this correlation. Using the prescribed methodology, the ITR of the model was determined to be 7.58 m$^2$K/GW, corresponding to an interfacial thermal conductance (ITC) of 131.93 MW/m$^2$K. To evaluate the reliability of the findings, Table 3 summarizes the ITR and ITC of various PMNCs obtained from previous numerical and experimental studies. It should be noted that there is no directly comparable data on the interfacial heat transfer in BNNS/PLA systems available in the literature. Therefore, the presented data includes different systems with varied matrices and fillers. Wang et al. [37] reported that the experimentally determined ITR of the BNNS/EVA film was 7.6 m$^2$K/GW, which supports the predicted value. Chaurasia et al. [86] investigated the ITC of BNNS/polyethylene nanocomposites with the assistance of MD simulations and estimated that ITC was 158.6±3.89 MW/m$^2$K, which is in agreement with the reported value of ITC in this study. MD results of BNNS/hexanol (7.3 m$^2$K/GW) and BNNS/hexanoic acid (6.4 m$^2$K/GW) also provide evidence for the determined value [87]. Furthermore, Yang et al. [88], studied the interfacial heat transfer at the BNNS/epoxy interface and reported an ITR of 3.62 m$^2$K/GW, which is in the same order as the value in the current work. It can be observed that the ITR/ITC values of PMNCs containing BNNS are similar to those reported in this study, which validates the reliability of the results. It is worth mentioning that the high surface area-to-volume ratio of BNNSs promotes strong molecular interactions with the polymer matrix, resulting in much lower ITR in BNNS-reinforced PMNCs compared to those embedded with bulk BN [24, 89]. Additionally, Bellussi et al. [90] demonstrated that utilizing graphene sheets within the PLA matrix would significantly enhance the ITR values. The improved interfacial heat transfer in the presence of BNNSs can be ascribed to the emergence of high-frequency phonon vibrations in these nanofillers, which will be further elaborated in Section 3.3.

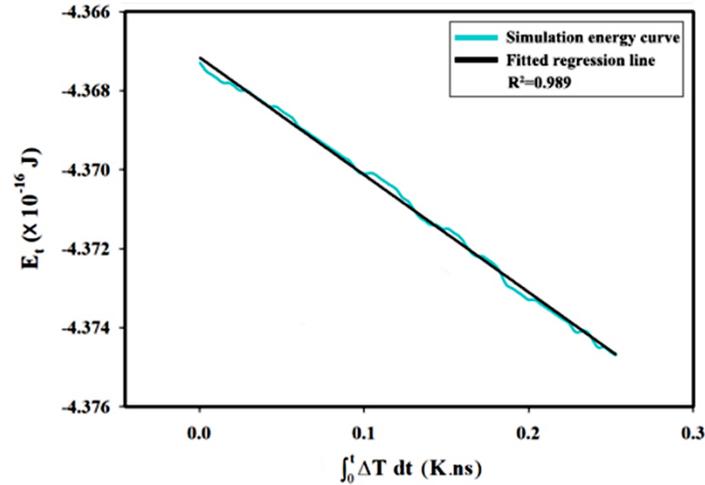

Fig. 6. Variations in BNNS total energy during interfacial heat transfer simulation time, correlated with the $T_{BNNS}$-$T_{PLA}$ integration. The linear fitting produces the interfacial thermal conductance directly.

Table 3. ITR and ITC of the BNNS/PLA model in the present study in comparison with the available data in the literature for various PMNCs.

| Reference | Sample | ITR [m²K/GW] | ITC [MW/m²K] |
|---|---|---|---|
| **Present work** | BNNS/PLA | 7.58±0.29 | 131.93 |
| [37] | BNNS/EVA | 7.6 | - |
| [88] | BNNS/Epoxy | 2.42-3.62 | - |
| [86] | BNNS/PE | - | 158.6±3.89 |
| [37] | BN/EVA | 24 | - |
| [87] | BN/hexanol | 7.3 | - |
| [87] | BN/hexanoic | 6.4 | - |
| [90] | G/PLA | 142 | - |
| [90] | GO/PLA | 88 | - |
| [90] | rGO/PLA | 75 | - |
| [91] | G/paraffin | 16.3 | - |
| [92] | G/Epoxy | 3.5 | - |

### 3.3. On the role of functional groups

The interfacial area of nanocomposites is typically characterized by the weak van der Waals interactions, resulting in high interfacial thermal resistance [93]. It has been well-established that the formation of covalent bonds at this region can significantly enhance the transfer of thermal energy across the interface. Functionalizing the surface of embedded nanofillers to enhance interfacial binding energy has proven to be an effective strategy for improving thermal transport in PMNCs [91, 94–96]. Consequently, the positive impact of functionalizing BNNS on the thermal properties of polymer matrices has been recognized. In this study, we investigated the role of surface treatment on the ITR of BNNS/PLA nanocomposites using three different functional groups: hydroxyl groups (–OH), amino groups (–NH$_2$), and carboxyl groups (–COOH). Through a series of systematic MD simulations, the effects of these groups were compared to analyze the influence of their polarity on the interfacial thermal behavior of the nanocomposite samples. Subsequently, we explored the role of the coverage degree and arrangement of these groups on the ITR of BNNS/PLA nanocomposites using the introduced pump-probe technique.

#### 3.3.1. Effect of coverage degree

To evaluate the impact of coverage degree on the ITR of the samples under study, we applied 3 different levels of coverage to the BNNS, with –OH molecules serving as the functionalizing group. This quantity represents the percentage of atoms bonded to functional groups [97]. As shown in Table 4, the ITR of nanocomposite models containing surface-functionalized BNNS at coverage values of 2.5%, 5% and 7.5% of hydroxyl group were found to be 6.37±0.32, 6.17±0.38, and 5.93±0.35 m$^2$K/GW, respectively. These values demonstrate a reduction of 15.9%, 18.6%, and 21.8%, respectively, compared to the ITR of the pristine BNNS/PLA model (i.e., 7.58±0.29 m$^2$K/GW). This reverse correlation was also observed in the models containing BNNS functionalized with –COOH and –NH$_2$ groups. Utilizing the same approach, a similar trend was reported by Bellussi et al. [90] for the ITR of graphene/PLA nanocomposites in the presence of hydroxyl molecules. Additionally, Liu et al. [98] found that the addition of –OH groups significantly enhances the ITC of BNNS-epoxy nanocomposites. To explore the underlying mechanism, extensive analysis of the vibrational density of states (VDOS) in various composite systems has been conducted [99, 100]. It has been demonstrated that introducing more chemical bonds into the interfacial area improves heat transfer efficiency and enhances thermal transfer across the interface through vibrational coupling between the composite constituents. Thus, in the OH-BNNS/PLA composite models examined in this study, the presence of more covalent bonds between functionalized BNNS and PLA would reduce phonon scattering at the interface due to interfacial compatibility and strong interfacial interactions, leading to improved thermal conductance of the samples. To further elucidate this issue, the VDOS analysis of the pristine PLA model and the nanocomposite samples infused with untreated and surface-functionalized BNNS has been provided in the following sections.

Table 4. ITR and ITC of BNNS/PLA nanocomposite models in the presence of –OH, –NH$_2$, and –COOH groups at various coverage degrees.

| Functionalized nanocomposite samples | ITR [m$^2$K/GW] | ITC [MW/m$^2$K] |
|---|---|---|
| BNNS/PLA | 7.58±0.29 | 131.93 |
| 2.5%OH-BNNS/PLA | 6.37±0.32 | 156.99 |
| 5%OH-BNNS/PLA | 6.17±0.38 | 162.07 |
| 7.5%OH-BNNS/PLA | 5.93±0.35 | 168.63 |
| 2.5%COOH-BNNS/PLA | 6.30 | 158.73 |
| 5%COOH-BNNS/PLA | 6.09 | 164.20 |
| 7.5%COOH-BNNS/PLA | 5.79 | 172.71 |
| 2.5%NH$_2$-BNNS/PLA | 6.65 | 150.38 |
| 5%NH$_2$-BNNS/PLA | 6.32 | 158.23 |
| 7.5%NH$_2$-BNNS/PLA | 6.00 | 166.67 |

### 3.3.2. *Influence of polarity*

After investigating the impact of coverage degree on the ITR of the BNNS/PLA models under study, we proceeded to examine the relationship between the interfacial thermal properties of the samples and the polarity of their functional groups (refer to Table 4). In general, the introduction of functional groups to pristine nanofillers, such as BNNS in this case, tends to enhance the attraction between the composite constituents. However, oxygen containing groups, namely –OH and –COOH, exhibited significantly stronger interfacial interactions compared to –NH$_2$. This can be attributed to their ability to enhance van der Waals interactions and facilitate the formation of new hydrogen bonds in the interfacial area [50]. Consequently, nanocomposites with more polarized functional groups demonstrated higher ITC, which correlates well with the increased attraction of BNNS towards the interface in these cases, owing to stronger electrostatic interactions. The shorter interatomic distance in samples with more polarized functional groups contributes significantly to the augmentation of heat flux through electrostatic and van der Waals interactions [87]. Therefore, at a constant coverage degree, carboxyl and hydroxyl groups exhibit the most effective enhancement of ITC in the aforementioned nanocomposites.

### 3.3.3. Towards the underlying mechanisms

To gain a deeper understanding of the mechanisms governing the influence of nanofiller surface-treatment on the ITR of BNNS/PLA nanocomposites, we focused on two analysis methods: the radial distribution function (RDF) and the VDOS. First, microstructural characterization of models containing 5% of $-OH$, $-COOH$, and $-NH_2$ molecules was conducted as one of the main fingerprints used to study on how the interfacial bonding strength can be affected by the presence of different functional groups. The RDF curves, depicted in Fig. 7, were calculated to facilitate a comparative analysis. Additionally, the interfacial properties of the sample infused with untreated BNNS are also presented in this figure. It is worth noting that the appearance of sharp peaks in the RDF indicates the formation of chemical bonds at the interface. These bonds act as bridges, tightly connecting BNNS and PLA, thereby facilitating enhanced heat transfer across this region.

Referring back to the results discussed in the previous section, the COOH-BNNS/PLA nanocomposite system exhibited the most favorable interfacial thermal features among all the models examined in this study. This observation is reflected in the trends observed in the RDF calculations shown in Fig. 7, where the peaks are more pronounced for the samples containing nanofillers coated with carboxyl groups. This suggests that the addition of $-COOH$ molecules increases the probability of forming ionic bonds at the interface, which is consistent with the reduction in ITR observed in these models.

To complete the discussion, we also sought to comprehend the interfacial thermal transport in the aforementioned nanocomposite samples by analyzing the VDOS, which refers to the number of vibrational modes per unit volume and frequency within the interfacial area. We obtained the VDOS by calculating the Fourier transform of the atomic velocity autocorrelation functions (VACF) as described in Refs [1, 82, 101]:

$$P(\omega) = \sum_i \frac{m_i}{k_B T} \int_0^\infty \left\langle v_i(t) \cdot v_i(0) \right\rangle exp(-i\omega t) \, dt \qquad (4)$$

where $m_i$, $v_i$, and $\omega$ are the mass and velocity of atom $i$, and the angular frequency, respectively. It is well-established that mismatches in the vibrational properties of materials in contact play a significant role in phonon transport across their interfaces [102]. When the VDOS of materials on both sides of an interface have a high degree of overlap, it indicates strong vibrational coupling between the two materials. As a result, phonon mismatches at the interface are reduced, leading to lower thermal resistance in this area. Therefore, by comparing the overlap between the VDOSs for specific frequency regions, we can reveal a better interface thermal transport between the constituents of a nanocomposite system. As shown in Fig. 8, the VDOS analysis of the neat PLA model reveals two sharp peaks at approximately 45 THz and 90 THz, which is consistent with results obtained by other researchers [90]. Based on the VDOS curves, there appears to be a noticeable mismatch between the BNNS/PLA and PLA models. However, when 7.5% of carboxyl functional groups are introduced on the nanofiller surface, the VDOS curve overlaps with that of PLA due to the coupling strength. As the overlapping areas increase, more channels are activated between phonon modes at the same frequency, indicating superior compatibility of the VDOS spectra at the interface of COOH-BNNS/PLA composites. This leads to enhanced heat transfer efficiency and decreased interfacial thermal resistance in these samples.

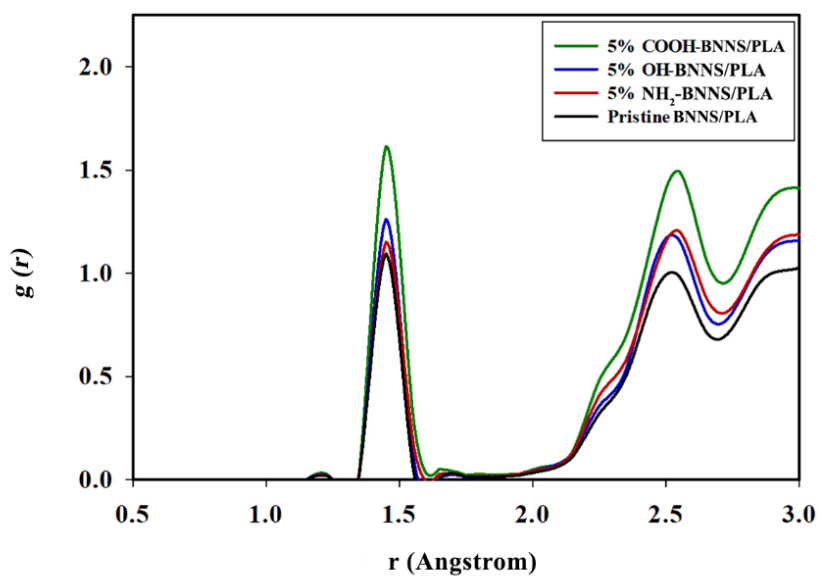

Fig. 7. RDF of BNNS/PLA models containing 5% COOH, OH, and NH$_2$ groups to investigate the formation of chemical bonds at the interface.

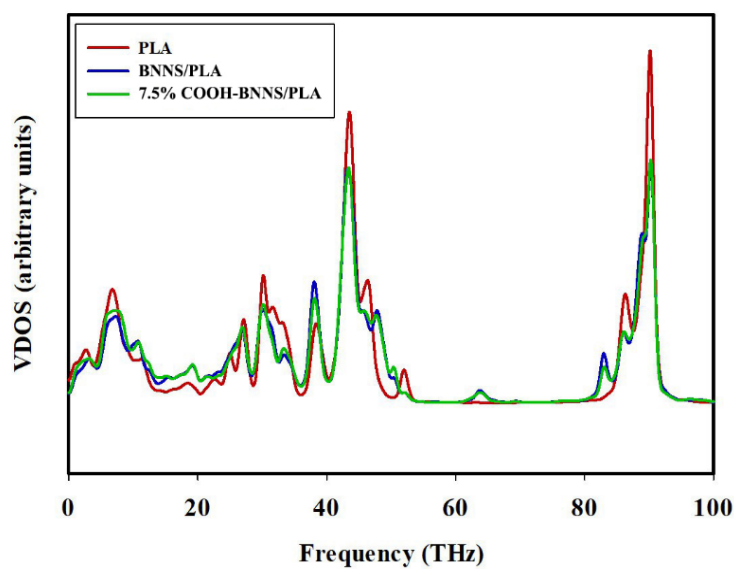

Fig. 8. VDOS of neat PLA, BNNS/PLA, and 7.5% COOH-BNNS/PLA models obtained based on the Fourier transform of VACF.

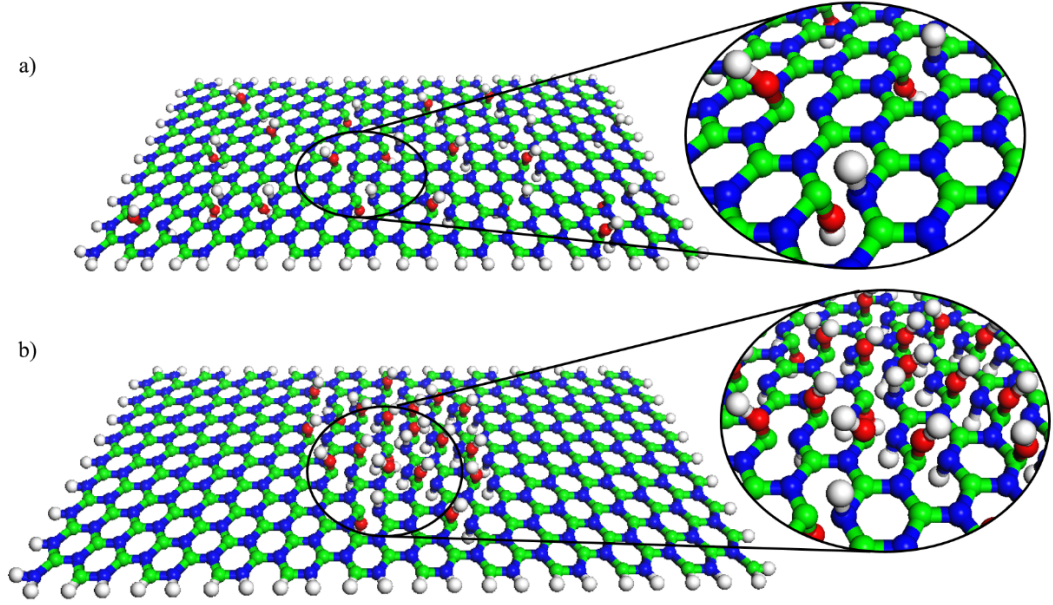

Fig. 9. Two different arrangements of 5% –OH groups: a) Random and b) Agglomerated pattern.

### 3.3.4. The Impact of functional group arrangement

Previous experimental and numerical studies have shown that, in addition to the type and coverage degree of functionalized groups, the mechanical and thermal properties of PMNCs embedded with surface-treated nanofillers are also influenced by the arrangement of functional molecules. In order to further explore this aspect, we conducted an investigation on the impact of different patterns of hydroxyl groups, deposited on BNNS at a coverage of 5%, and their effect on the ITR between the nanofiller and PLA. As illustrated in Fig. 9, two distinct arrangements of –OH groups were considered in this study: random and aggregated patterns. For the model with randomly distributed –OH groups, an ITR value of 6.17±0.38 m²K/GW was obtained, as reported in Table 4. On the other hand, in the case of the sample reinforced with 5% agglomerated –OH groups, the ITR was estimated to be 6.24±0.37 m²K/GW. These results indicate that the nanocomposite model with agglomerated hydroxyl molecules experienced a further decline in the ITC value. This arises because the mentioned procedure converts the local B–N bonds from sp² to sp³ hybridization, which promotes phonon scattering [103]. Accordingly, the agglomerated functionalization exacerbates this issue by introducing a new interface for phonon scattering. Additionally, the weaker sp³ hybridizations in the agglomerated –OH groups allow for rotation as heat is transported through the sheet. Consequently, localized sp³-bonding phonons can interact with other low-energy phonons, leading to a decrease in their mean free path [103]. These findings suggest that the system containing agglomerated functional groups exhibits larger ITR, indicating a reduction in the interfacial thermal transport.

## 4. Conclusion

In summary, although PLA exhibits promising physical and mechanical properties, its limited thermal conductivity restricts its widespread application. To overcome this inherent limitation, nanostructures have been incorporated into PLA-based nanocomposites as a second phase. This study systematically examined the impact of embedding pristine and surface-treated BNNSs on the ITR of PLA-based nanocomposites through a series of MD simulations. The key findings can be summarized as follows:

- The addition of 3 wt.% BNNS to PLA resulted in a nanocomposite ITR of 7.58±0.29 $m^2K/GW$, which is consistent with previous similar studies.

- Functionalized nanosheets significantly affected the ITR of nanocomposites. Functional groups effectively enhanced the interactions between BNNS and PLA, improved vibrational coupling, reduced phonon mismatch at the interface, and thereby enhanced cooling efficiency.

- The degree of functionalization played a crucial role in heat transfer across the interface. With increased coverage of functional groups, the ITR decreased in a subtractive manner. This reduction was attributed to the decrease in phonon mismatch at the interface as the ratio of functional groups increased.

- Covalently functionalized BNNSs with –OH, –$NH_2$, and –COOH groups were evaluated for their ITR. The stronger electrostatic interactions resulting from more polarized functional groups attracted BNNS toward the interface, ultimately enhancing interfacial thermal transport. These findings were supported by various analyses, including RDF calculation and VDOS analysis.

- Oxygen-containing groups (–OH and –COOH) exhibited significantly better interfacial interactions than –$NH_2$. This can be attributed to their ability to promote van der Waals interactions and form new hydrogen bonds.

- A slight tuning in the surface modification of nanofillers (2.5, 5, and 7.5%) is sufficient to significantly improve the interfacial thermal conductance of nanocomposites. In the best-case scenario a 31% improvement in thermal transport for COOH-BNNS/PLA with a 7.5% functionalization degree was observed.

- The thermal properties of the interface were slightly improved when functional groups were randomly distributed compared to an agglomerated pattern.
- These findings provide valuable insights into selecting functional groups for nanofillers to enhance the thermal performance of PMNCs used in thermal management systems and biomaterials.

This study also presents an effective approach to increasing the thermal conductivity of PLA by incorporating BNNSs, as well as a deeper understanding of the thermal mechanisms at the interface. While these findings are promising, further research is necessary to determine the optimal degree of functionalization. Additionally, future investigations could explore how the ITR of BNNS/PLA nanocomposites responds to temperature and system dimensional variations. This could be achieved using other MD approaches such as steady-state non-equilibrium molecular dynamics (SNEMD).

**Declaration of Competing Interest**

The authors declare that there is no conflict of interests regarding the publication of this article.

**Data Availability**

The raw/processed data required to reproduce these findings cannot be shared at this time as the data also forms part of an ongoing study.